\documentclass[aps,prd,twocolumn,groupedaddress,showpacs,showkeys]{revtex4}
\usepackage{graphicx}% Include figure files
\usepackage{dcolumn}% Align table columns on decimal point
\usepackage{bm}% bold math
\usepackage{amssymb,amsmath,latexsym,amsfonts}
\begin{document}

%\preprint{APS/123-QED}

\title{Critical Temperature Associated to Symmetry Breaking of Klein--Gordon fields
versus Condensation Temperature in a Weakly interacting Bose--Einstein Gas}% Force line breaks with \\

\author{El\'ias Castellanos}
 \email{ecastellanos@fis.cinvestav.mx} \affiliation{Departamento de F\'{\i}sica,
 Centro de Investigaci\'on y Estudios Avanzados del IPN\\
 A. P. 14--740,  07000, M\'exico, D.F., M\'exico.}

\author{Tonatiuh Matos\footnote{Part of the Instituto Avanzado de Cosmolog\'ia (IAC) collaboration http://www.iac.edu.mx/}}
 \email{tmatos@fis.cinvestav.mx}\affiliation{Departamento de F\'{\i}sica,
 Centro de Investigaci\'on y Estudios Avanzados del IPN\\
 A. P. 14--740,  07000, M\'exico, D.F., M\'exico.}

\date{\today}% It is always \today, today,

\begin{abstract}
We deduce the relation between the critical temperature associated
to the symmetry breaking of  scalar fields with one--loop correction
potential immersed in a thermal bath and the condensation
temperature of the aforementioned system, assuming a harmonic
oscillator type potential. We show that these two temperatures are
related through the \emph{scale} associated to the system. In this
aim, we infer the order of magnitude for the \emph{scale} as a
function of the corresponding healing length, in order to give a
criterium to compare both temperatures. Additionally, we prove that
the condensation temperature is independent of the thermal bath
within the semiclassical approximation, for a positive coupling
constant, assuming that the thermal bath contribution is the lowest
energy associated to the system.
\end{abstract}

\maketitle

%****************************************************************************
\section{Introduction}

Since its observation with the help of magnetic traps
\cite{Anderson}, the phenomenon of Bose--Einstein condensation has
spurred an enormous amount of works on the theoretical and
experimental realms associated to this topic. The principal interest
in the study on Bose--Einstein condensation is its interdisciplinary
nature. From the thermodynamic point of view, this phenomenon can be
interpreted as a phase transition, and from the quantum mechanical
point of view as a matter wave coherence arising from overlapping de
Broglie waves of the atoms, in which many of them condense to the
grown state of the system. In quantum field theory, this phenomenon
is related to the spontaneous  breaking of a gauge symmetry
\cite{Bagnato}. Symmetry breaking is one of the most essential
concepts in particle theory and has been extensively used in the
study of the behavior of particle interactions in many theories
\cite{Pinto}. The concept with the accompanying wave function
describing the condensate, was first introduced in explaining
superconductivity and super fluidity \cite{Griffin}. Phase
transitions are changes of state, related with changes of symmetries
of the system. The analysis of Symmetry breaking mechanisms have
turn out to be very helpful in the study of phenomena associated to
phase transitions in almost all areas of physics. Bose-Einstein
Condensation is one topic of interest that uses in an extensive way
the Symmetry breaking mechanisms \cite{Bagnato}, its phase
transition associated with the condensation of atoms in the state of
lowest energy and is the consequence of quantum, statistical and
thermodynamical effects.

On the other hand, the results from finite temperature quantum field
theory \cite{Dolan,Weinberg} raise important challenges about their
possible manifestation in condensed matter systems. By investigating
the massive Klein--Gordon equation (KG), in \cite{Matos:2011kn} we
were able to show, that the Klein--Gordon equation can simulate a
condensed matter system. In \cite{Matos:2011kn} it was shown that
the Klein--Gordon equation with a self interacting scalar field (SF)
in a thermal bath reduces to the Gross--Pitaevskii equation in the
no-relativistic limit, provided that the temperature of the thermal
bath is zero. Thus, the Klein--Gordon equation reduces to a
generalized relativistic, Gross--Pitaevskii equation at finite
temperature. But a question remains open. The Klein--Gordon equation
with a self interacting scalar field potential defines a symmetry
breaking temperature, at which the system experiments a phase
transition. However, this phase transition does not necessarily
means a condensation of the particles of the system. Here, our aim
is to study a model made up of a real scalar field together with the
possibility that this SF might undergo a phase transition as the
temperature of the system is lowered. Nevertheless, the fact that
the system breaks its original symmetry at some temperature is not a
guarantee that the system undergoes a condensation at the same one.
The main aim of the present work is to obtain the temperature of
condensation of a SF system of particles and compare it with the
symmetry breaking temperature, using realistic systems. Consider the
easiest case of a double-well interacting potential for a real
scalar field $\Phi(\vec{x},t)$ that goes as

\begin{equation}
\label{Pot0}
V(\Phi)=-\frac{m^{2}c^{2}}{2\hbar^{2}}\Phi^{2}+\frac{\lambda}{4\hbar^{2}c^{2}}\Phi^{4}.
\end{equation}

Here one important idea, to which we shall refer, is that of
identifying the order parameter  which characterizes the phase
transition with the value of the real scalar quantum field $\Phi$.
From quantum field theory we know that the dynamics of a SF is
governed by the Klein-Gordon equation, it is the equation of motion
of a field composed of spinless particles. In this case we will add
an external field that will interact with the SF to first order, such
that the KG equation will be given by

\begin{equation}
\label{KG} \Box^2\Phi - \frac{d V(\Phi )}{d
\Phi}-2\frac{m^{2}c^{2}}{\hbar^{2}} \Phi \phi=0.
\end{equation}
where the D' Alambertian $\Box^2$ is defined as
\begin{equation}
\label{eq:DAlambertian}
\Box^2=\nabla^2-\frac{1}{c^{2}}\frac{\partial^{2}}{\partial t^{2}},
\end{equation}
and the external potential here is denoted by $\phi$. We suppose
that the scalar field has a temperature $T$ such that we can
consider that the scalar field is in a thermal bath of temperature
$T$. In this case, the scalar field can be described by the
potential (\ref{Pot0}) up to one loop in corrections. The scalar
field potential in a thermal bath at temperature $T$ is given by
\cite{Dolan,Weinberg}
\begin{eqnarray}
\label{Pot}
V(\Phi)&=&-\frac{m^{2}c^{2}}{2\hbar^{2}}\Phi^{2}+\frac{\lambda}{4\hbar^{2}c^{2}}\Phi^{4}
+\frac{\lambda}{8\hbar^{2}c^{2}} k_{B}^{2} T^{2}
\Phi^{2}\\\nonumber&-&\frac{\pi^{2}}{90 \hbar^{2}c^{2}}k_{B}^{4}
T^{4},
\end{eqnarray}
where  $k_{B}$ is the Boltzmann's constant, and $\lambda$ is the
self-coupling constant.

It is convenient to consider the total potential $V_{T}$ adding the
external one; using this fact, we are able to express (\ref{KG}) as
follows

\begin{equation}
\label{KG1} \Box^2\Phi + \frac{d V_{T}(\Phi )}{d \Phi}=0,
\end{equation}
where
\begin{eqnarray}
\label{Pot1}
V_{T}(\Phi)&=&-\frac{m^{2}c^{2}}{2\hbar^{2}}\Phi^{2}+\frac{\lambda}{4\hbar^{2}c^{2}}\Phi^{4}
+\frac{\lambda}{8\hbar^{2}c^{2}} k_{B}^{2} T^{2}
\Phi^{2}\\\nonumber&-&\frac{\pi^{2}}{90 \hbar^{2}c^{2}}k_{B}^{4}
T^{4}-\frac{m^2c^2}{\hbar^2}\phi{\Phi^2}.
\end{eqnarray}
When the temperature T is high enough, one of the minimums of the
potential (\ref{Pot}) is $\Phi=0$. On the other hand, when the
temperature is sufficiently small, the term proportional to $T^{4}$
is not longer important, as for sufficiently low $T$ the term that
goes as $T^{4}$ can be dropped out  (see \cite{Matos:2011kn}).

%****************************************************************************
\section{Critical Temperature from the Symmetry Breaking of Klein--Gordon Fields}
%\bigskip
%\bigskip
Let us calculate the critical temperature associated to the
Klein--Gordon equation \eqref{KG1}, with the corresponding potential
extended to one loop and immersed in a thermal bath \eqref{Pot1}.
Neglecting the term proportional to $T ^{4}$, assuming that the
temperature is sufficiently small, then, the critical temperature
where the minimum of the potential $\Phi=0$ becomes a maximum is
\cite{Matos:2011pd}

\begin{equation}
\label{TCS0}
k_{B}T_{c}^{SB}=\frac{2mc^{2}}{\sqrt{\lambda}}\Bigl(1+2\phi\Bigr)^{1/2}.
\end{equation}
This is the temperature at which the symmetry of
the system is broken. The fact that the field $\Phi$ is a real
scalar field, has as a consequence that the associated Lagrangian is
invariant under the transformation $\Phi\rightarrow-\Phi$ (this is
the so--called $Z_{2}$ symmetry, which is a discrete symmetry) \cite{Dehnen}, and
the Klein-Gordon equation (\ref{KG1}), is also invariant under this
symmetry. In this case, for temperatures $T>T^{SB}_{c}$ we have a
minimum in $\Phi=0$ for the potential (\ref{Pot1}), which is
invariant under the symmetry $\Phi\rightarrow-\Phi$. For
temperatures $T<T^{SB}_{c}$ we have two minima in
\begin{equation}
\Phi_{min}=\pm
\frac{mc^{2}}{\sqrt{\lambda}}\Bigg(1+2\phi-\lambda\Bigl(\frac{\kappa_{B}T}{2mc^{2}}\Bigr)^{2}\Bigg)^{1/2},
\label{min}
\end{equation}
these two minima are not invariant under the symmetry
$\Phi\rightarrow-\Phi$ (in fact it maps each of the two vacua into
the other). In this case, the symmetry is said to be spontaneously
broken \cite{kaku}. Expression (\ref{min}), allows us define the critical
temperature (\ref{TCS0}) at which the symmetry
$\Phi\rightarrow-\Phi$ of the two the minima of the potential
(\ref{Pot1}) is broken. However, we must add that the $T^{SB}_{c}$
is valid only close to the minimum $\Phi=0$.

Let us suppose that the external potential $\phi$ is time
independent and that can be described as a harmonic oscillator
type--potential $\phi \sim r^{2}$. Clearly, this can be extended to
a more general potentials. With these assumptions, the critical
temperature $T_{c}^{SB}$ near to the center of the system
($\vec{r}=0$)  is given by

 \begin{equation}
\label{TCS} k_{B}T_{c}^{SB}=\frac{2mc^{2}}{\sqrt{\lambda}}.
\end{equation}
We notice also, that in the case when $\lambda\rightarrow0$, the
critical temperature $T_{c}^{SB}\rightarrow\infty$.

%%%%%%%%%%%%%%%%%%%%%%%%%%%%%%%

In order to show how the Klein--Gordon equation is a generalization
of the Gross--Pitaevskii equation, we perform the transformation
 \begin{equation}
  {\kappa}\Phi=\frac{1}{\sqrt{2}}\left(\Psi\,\mathrm{e}^{-\mathrm{i}\hat mct}+\Psi^*\,\mathrm{e}^{\mathrm{i}\hat mct}\right)\label{eq:Phi}
 \end{equation}
where $\kappa$ is the scale of the system, which is to be determined
by an experiment, and $\hat{m}^2=m^2c^2/\hbar^2$.

In terms of function $\Psi$, the Klein--Gordon equation (\ref{KG})
reduces to,
\begin{eqnarray}
 \mathrm{i}\hbar\dot{\Psi}+\frac{\hbar^2}{2m}\Box^2{\Psi}+\frac{3\lambda}{2mc^2}|\Psi|^2\Psi-mc^2\phi\Psi
 +\frac{\lambda k_B^2T^2}{8mc^2}\Psi=0,\nonumber\\
\label{eq:Schrodinger0}
\end{eqnarray}
and an equivalent equation for the complex conjugate. Here
$\dot{}=\partial/\partial t$ and $|\Psi|^2=\Psi\Psi^*=\rho$. Let us
to remark that (\ref{eq:Schrodinger0}) is the Klein--Gordon equation
rewritten in terms of the function $\Psi$.
However, the transformation (\ref{eq:Phi}) is not unique.
For instance, we can change $\Psi$ to
$\Psi'=\Psi+\mathrm{i}Fe^{\mathrm{i}\hat{m}ct}$, with $F$ a real
field. If one inserts this new transformation in (\ref{eq:Phi})
then, $\Phi$ is unchanged and therefore $F$ is apparently
undetermined. Nevertheless, if we express $\Psi'$ in terms of a
modulus $n'$ and a phase $S'$ as
\begin{equation}
 \kappa \Psi'=\sqrt{ n'}\,\mathrm{e}^{\mathrm{i}S'}.
 \label{eq:psi}
 \end{equation}
then, we obtain an expression for the particle density $n'$
\begin{equation}
n'=\kappa^{2}\Big|\Psi'\Big|^{2} \label{n0},
\end{equation}
where $\kappa$ is the scale associated to the system. From (\ref{n0}), we obtain
an expression for the density associated to $\Psi'$ in function of $\Big|\Psi\Big|^{2}$ and $F^{2}$ given by
\begin{equation}
n'=\kappa^{2}\Bigl(\Big|\Psi\Big|^{2}+F^{2}+iF\,Im\{e^{\mathrm{i}\hat{m}ct}\Psi\}\Bigr),
\end{equation}
which in terms of the corresponding densities can be re-written as,
\begin{equation}
n'=n+n_{F}+J,
\end{equation}
where $J=\sqrt{n}\kappa F\sin(\hat{m}ct-S)$, being
$Im\{e^{\mathrm{i}\hat{m}ct}\Psi\}$ the imaginary part of $e^{\mathrm{i}\hat{m}ct}\Psi$. If we assume that $n$ and
$n_{F}$ are particle densities associated to the fields $\Big|\Psi\Big|^{2}$
and $F^{2}$ respectively, then we are able to interpreted the term
$J=\sqrt{n\,n_F} \sin(\hat{m}ct-S)$ as an oscillation of the particle density between these two
fields, provided that $\Big|\Psi'\Big|^{2}$ fullfils the
normalization condition over the total number of particles
\begin{equation}
 N=\int d^{3}\vec{r}\hspace{0.1cm}\Big|\Psi'\Big|^{2}.
\label{eq:N}
\end{equation}
Using these facts, we are able to determine the field $F$, resorting to equation (\ref{n0}) or (\ref{eq:N}).
On the other hand, observe that when $T=0$ and in the
non-relativistic limit, $\Box^2\rightarrow\nabla^2$, eq.
(\ref{eq:Schrodinger0}) becomes the Gross-Pitaevskii equation for
Bose-Einstein Condensates, which is an approximate equation for the
mean-field order parameter in the classical theory \cite{b.n}.
Equation (\ref{eq:Schrodinger0}) can also be written as
\begin{eqnarray}
 \mathrm{i}\hbar\dot{\Psi}+\frac{\hbar^2}{2m}\Box^2{\Psi}+\frac{3\lambda}{2mc^2}|\Psi|^2\Psi&-&mc^2\phi\Psi\nonumber\\
 &+&\frac{mc^2}{2}\left(\frac{T}{T_{c}^{SB}}\right)^2\Psi=0,\nonumber\\
\label{eq:Schrodinger}
\end{eqnarray}
which implies that if $T<<T_{c}^{SB}$, the last term of equation
(\ref{eq:Schrodinger}) can be neglected. We expect that this
equation is a good approximation for the condensation process close
to the symmetry breaking temperature $T_{c}^{SB}$. The static limit
of equation (\ref{eq:Schrodinger0}) or (\ref{eq:Schrodinger}) is
known as the Ginzburg-Landau equation. This is the reason why we
interpret (\ref{eq:Schrodinger0}) as a generalization of the
Gross-Pitaevskii equation for finite temperatures and relativistic
particles (see \cite{Matos:2011pd} for details).
%%%%%%%%%%%%%%%%%%%%%%%%%%%%%%%
Nevertheless, the critical temperature for the break down of the
symmetry is not necessarily a sign of condensation, $T_c^{SB}$ could
be different to the critical temperature of condensation.  The main
goal of this work is to compare both temperatures and to give some
ideas how to compare them with experiments in the laboratory.

\section{Condensation Temperature}

In order to calculate the condensation temperature associated to the
aforementioned system and its relation with the critical temperature
$T^{SB}_{c}$, let us insert plane waves in equation (\ref{KG}). This
fact allows us to obtain the single--particle dispersion relation
between energy and momentum  \cite{ZWIE}, which we interpreted as
the semiclassical energy spectrum associated to the Klein--Gordon
equation in a thermal bath, with the result (we considered here the
low velocities limit)

\begin{equation}
\label{ES0} E_{p}\simeq
mc^{2}+\frac{p^{2}}{2m}+\frac{\lambda}{2mc^{2}}\Big|\Phi
\Big|^{2}+\frac{\lambda}{4mc^{2}}(\kappa_B T)^{2}+mc^{2} \phi
\end{equation}

Additionally, we interpreted $\kappa^{2}\Big|\Phi(\vec{r},t)
\Big|^{2}$ as the spatial density $n(\vec{r},t)$ of the cloud, where
$\kappa$ is the scale of the system. Assuming static thermal
equilibrium $n(\vec{r},t)\approx n(\vec{r})$ \cite{Dalfovoro}, then

\begin{equation}
\label{FE} \Big|\Phi \Big|^{2}\equiv\kappa^{-2}n(\vec{r}).
\end{equation}

Using (\ref{FE}) and neglecting the rest mass, we can re--write the
semiclassical energy spectrum (\ref{ES0}) as follows

\begin{equation}
\label{ES}
E_{p}\simeq\frac{p^{2}}{2m}+\frac{\lambda\kappa^{-2}}{2mc^{2}}
n(\vec{r})+\frac{\lambda}{4mc^{2}} (\kappa_B T)^{2}+mc^{2} \phi.
\end{equation}

Is noteworthy to mention that if we set
$\lambda\equiv16\pi\hbar^{2}c^{2}\kappa^{2}a$ and $\phi\equiv\alpha
r^{2}$, with $\alpha\equiv1/2(\omega_{0}/c)^{2}$, where $a$ is the
s--wave scattering length and $\omega_{0}$ is the frequency of an
isotropic harmonic oscillator say (this can be generalized for
different types of potentials), we obtain the semiclassical energy
spectrum in the Hartree--Fock approximation for a bosonic gas
trapped in an external potential (an isotropic harmonic oscillator
according to our previous assumptions), but with an extra term due
to the contributions of the thermal bath. The Hartree--Fock
approximation consists basically in the assumption that the
constituents of the gas behave like a non--interacting bosons,
moving in a self--consistent mean field and is valid when
$E_{p}>>\mu$, being $\mu$ the chemical potential
\cite{Dalfovoro,Pethick}. On the other hand, within the
semiclassical approximation, the spatial density $n(\vec{r})$ reads
\cite{Dalfovoro,Pethick}

\begin{equation}
\label{DE} n(\vec{r})=\frac{1}{(2 \pi \hbar)^{3}}\int  d^{3} \vec{p}
\hspace{0.1cm} n(\vec{r},\vec{p}),
\end{equation}
where $n(\vec{r},\vec{p})$ is the Bose--Einstein distribution
function given by \cite{Dalfovoro,Pethick}
\begin{equation}
\label{Sd}
 n(\vec{r},\vec{p})=\frac{1}{e^{\beta(E_{p}-\mu)}-1}.
\end{equation}
where $\mu$ is the chemical potential, $\beta=1/k_{B} T$, and
$E_{p}$ is the semiclassical energy spectrum. The number of
particles in the 3--dimensional space obey the normalization
condition \cite{Dalfovoro,Pethick}
\begin{equation}
\label{NC} N=\int d^{3}\vec{r}\hspace{0.1cm}n(\vec{r}).
\end{equation}
Let us analyze expression (\ref{NC}). From the grand
canonical ensemble formalism, one can derives the Bose-Einstein
distribution function (\ref{Sd}) which is valid for bosons
\cite{Greiner,Phatria}. In this formalism the number of particles is
a fixed quantity given by
\begin{equation}
N=\sum_{E}\frac{1}{e^{\beta(E-\mu)}-1},
\end{equation}
where $E$ is the single particle energy. Resorting to the concept of
thermodynamic limit \cite{yukalov}, we can replace the sum by an
integral, this is the so-called continuum approximation
\cite{Greiner,Phatria},
\begin{equation}
\label{NCC}
 N=N_{0}+\frac{1}{(2\pi\hbar)^{3}}\int_{0}^{\infty} \Omega(E)n(\vec{r},\vec{p})dE,
\end{equation}
where $N_{0}$ is the number of particles in the ground state and
$\Omega(E)$ is the number of microstates per energy unit
\cite{Greiner,Phatria}. Expression (\ref{NCC}) is totally equivalent
to (\ref{NC}). In the grand canonical ensemble, the total number of
particles and the total energy associated to the system in question
are fixed \cite{Greiner,Phatria}, which means, a conserved
quantities. Since the chemical potential $\mu$ is not fixed, but the
total number of particles $N$  and the total energy associated to
the system are conserved \cite{Phatria,Greiner} then, the value of
the chemical potential $\mu$ has to be determined using the
constraint that the number of particles is fixed.
Let us calculate the condensation temperature. Integrating the
spatial density (\ref{DE}) in the momentum space, using the
semiclassical energy spectrum (\ref{ES}) associated with the KG
equation (\ref{KG}), together with the Bose--Einstein distribution
function (\ref{Sd}) leads us to
\begin{equation}
\label{DE1}
n(\vec{r})=\Bigg(\frac{mk_{B}T}{2\pi\hbar^{2}}\Bigg)^{3/2}g_{3/2}(z(\vec{r})),
\end{equation}
where $z(\vec{r})$ is given by
\begin{equation}
\label{F}
z(\vec{r})=e^{\beta(\mu-\frac{\lambda\kappa^{-2}}{2mc^{2}}n(\vec{r})-\frac{\lambda}{4mc^{2}}(\kappa_B
T)^{2}-mc^{2}\phi)}.
\end{equation}
The function $g_{\nu}(z)$ is the so--called Bose--Einstein function
defined by \cite{Pethick,Phatria}
\begin{equation}
\label{BEF}
g_{\nu}(z)=\frac{1}{\Gamma(\nu)}\int_{0}^{\infty}\frac{x^{\nu- 1}
dx}{z^{-1} e^{x}-1}.
\end{equation}
With the use of the harmonic oscillator--type--potential
$\phi=\alpha r^{2}$, the function $z(\vec{r})$ can be re--written as
follows

\begin{equation}
\label{F1}
z(\vec{r})=e^{\beta(\mu-\frac{\lambda\kappa^{-2}}{2mc^{2}}n(\vec{r})-\frac{\lambda}{4mc^{2}}(\kappa_B
T)^{2}-mc^{2}\alpha r^{2})}.
\end{equation}
If $\lambda=0$ and $r=0$, $z(\vec{r})$ is just the fugacity
$z=\exp(\beta \mu)$ \cite{Pethick,Phatria}.

Expanding (\ref{DE1}) at first order in the coupling constant
$\lambda$, using the properties of the Bose--Einstein functions,
\cite{Phatria}
\begin{equation}
\label{BEF1}
 x\frac{\partial}{\partial x}g_{\nu}(x)=g_{\nu-1}(x),
\end{equation}
allows us to express the spatial density as follows

\begin{eqnarray}
\label{DE2} n(\vec{r})\approx n_{0}(\vec{r})&-&\lambda\,\,\,
g_{1/2}\Bigl(e^{\beta(\mu-\alpha
mc^{2}r^{2})}\Bigr)\\\nonumber&\times&\Bigg[\Bigl(\frac{k_{B}T}{2\kappa
c}\Bigr)^{2} \Bigl(\frac{m}{\pi\hbar^{2}}\Bigr)^{3}
g_{3/2}\Bigl(e^{\beta(\mu-\alpha mc^{2}r^{2})}\Bigr)\\\nonumber
&-&\sqrt{\frac{k_{B}^5
T^5m}{2^{5}\pi^{3}\hbar^{6}c^{4}}}\frac{g_{3/2}\Bigl(e^{\beta(\mu-\alpha
mc^{2}r^{2})}\Bigr)}{g_{1/2}\Bigl(e^{\beta(\mu-\alpha
mc^{2}r^{2})}\Bigr)}\Bigg],
\end{eqnarray}
where
\begin{equation}
\label{DE3}
n_{0}(\vec{r})=\Bigg(\frac{mk_{B}T}{2\pi\hbar^{2}}\Bigg)^{3/2}g_{3/2}\Bigl(e^{\beta(\mu-\alpha
mc^{2}r^{2})}\Bigr),
\end{equation}
is the density for case $\lambda=0$.
Integrating the normalization condition (\ref{NC}), using
(\ref{DE2}) lead us to
\begin{eqnarray}
\label{NC1}
N&\simeq&\frac{1}{\sqrt{2\alpha^{3}}}\Bigl(\frac{k_{B}T}{\hbar
c}\Bigr)^{3}g_{3}(e^{\beta \mu})\\\nonumber&-&\lambda\Bigg[
\frac{\kappa^{-2}}{8\hbar^{6}c^{5}}\Bigl(\frac{m}{\pi
\alpha}\Bigr)^{3/2}G_{3/2}(e^{\beta \mu})(k_{B}T)^{7/2}\\\nonumber
&-&\frac{1}{(2\alpha)^{3/2}m \hbar^{3}c^{5}}g_{2}(e^{\beta
\mu})(k_{B}T)^{4}\Bigg],
\end{eqnarray}
where

\begin{equation}
 G_{3/2}(z)=\sum_{i,j=1}^{\infty}\frac{z^{(i+j)}}{i^{1/2}j^{3/2}(i+j)^{3/2}},
\end{equation}
being $z=\exp(\beta \mu)$ the fugacity. If we set $\lambda=0$ in
expression (\ref{NC1}), we may obtain the expression for the number
of particles for the non--interacting case. At the condensation
temperature in the thermodynamic limit in the case $\lambda=0$,
$\mu=0$ \cite{Pethick}, and assuming that the number of particles in
the ground state above the condensation temperature is negligible,
allows us obtain an expression for the condensation temperature for
the non--interacting case $T_{0}$, given by
\begin{equation}
\label{CTI} k_{B}T_{0}=\Bigl(\frac{N\sqrt{2\alpha^{3}}}{\zeta
(3)}\Bigr)^{1/3}\hbar c,
\end{equation}
where $\zeta(x)$ is the Riemann Zeta function. 
Additionally, setting $\alpha\equiv1/2(\omega_{0}/c)^{2}$, we recover the  condensation
temperature for a bosonic gas trapped in an isotropic harmonic
oscillator potential \cite{Pethick}.
In order to obtain the leading correction in the shift for the
critical temperature respect to $T_{0}$ caused by the coupling
constant $\lambda$ and the thermal bath in our bosonic gas, let us
expand the expression (\ref{NC1}) at first order in $T=T_{0}$,
$\mu=0$, and $\lambda=0$, $T_{0}$ is the condensation temperature
for the non--interacting case (\ref{CTI}), with the result

\begin{eqnarray}
\label{NC2} N&=&\frac{1}{\sqrt{2\alpha^{3}}(\hbar
c)^{3}}\Bigg[\zeta(3)(k_{B}T_{0})^{3}
\\\nonumber&+&3\zeta(3)(k_{B}T_{0})^{2}k_{B}[T-T_{0}] +(k_{B}
T_{0})^{2}\zeta(2)\mu \Bigg] \\\nonumber&-&\lambda
\Bigg[\frac{\kappa^{-2}}{8\hbar^{6} c^{5}}\Bigl(\frac{m}{\pi
\alpha}\Bigr)^{3/2}
G_{3/2}(1)(k_{B}T_{0})^{7/2}-\frac{\zeta(2)(k_{B}T_{0})^{4}}{(2\alpha)^{3/2}
m \hbar^{3} c^{5}} \Bigg].
\end{eqnarray}
At the condensation temperature the chemical potential in the
semiclassical approximation is given by
\begin{equation}
\label{PQ}
\mu_{c}=\frac{\lambda\kappa^{-2}}{2mc^{2}}n(\vec{r}=0)+\frac{\lambda}{4mc^{2}}(k_{B}T_{c})^{2},
\end{equation}
as it is suggested from expressions (\ref{DE1}) and (\ref{F1}).
Expression \eqref{PQ} basically corresponds to the definition of the
chemical potential at the condensation temperature in the usual case
\cite{Dalfovoro}, except for the extra term contribution due to the
thermal bath.
Inserting (\ref{PQ}) in (\ref{NC2}) at the condensation temperature,
this allows us to obtain the shift caused by $\lambda$  in the
condensation temperature for the corresponding system, in function
of the number of particles

\begin{equation}
\label{SHIFT} \frac{T_{c}-T_{0}}{T_{0}}\equiv
 \frac{\Delta
T_{c}}{T_{0}}\simeq-\lambda\kappa^{-2}\alpha^{1/4}\frac{m^{1/2}}{c^{3/2}\hbar^{5/2}}\Theta
N^{1/6}.
\end{equation}
where

\begin{equation}
\label{CTE}
\Theta=\Bigg(\frac{\zeta(3)\zeta(2)-G_{3/2}(1)}{3(2\pi)^{5/4}\zeta(3)}\Bigg)
\Bigg(\frac{4}{\pi^{3}\zeta(3)}\Bigg)^{1/6}.
\end{equation}

Surprisingly, the thermal bath does not contribute to the shift in
the condensation temperature, because of the definition for the
chemical potential given in expression (\ref{PQ}) at the
condensation temperature. 
Analyzing the critical temperature caused by the symmetry breaking
expression (\ref{TCS}), with respect to the shift in the
condensation temperature expression (\ref{SHIFT}), assuming from the
very beginning that $\lambda>0$, we notice that when
$\lambda\rightarrow0$, then $T^{SB}_{c}\rightarrow\infty$ and the
shift $\Delta T_{c}/T_{0}\rightarrow0$, which means $T_{c}\simeq
T_{0}$. This implies that $T^{SB}_{c}>T_{0}>T_{c}$.

From expression (\ref{TCS}) and (\ref{SHIFT}), we obtain a relation
between $T^{SB}_{c}$ and $\Delta T_{c}/T_{0}$ given by

\begin{equation}
\label{REL} T^{SB}_{c}=\Bigl(1-T_{r}\Bigr)^{-1}
\frac{\kappa^{-1}\alpha^{1/8}}{k_{B}}\Bigl(\frac{mc}{2\pi\hbar}\Bigr)^{5/4}2\Theta
N^{1/12}.
\end{equation}
where $T_{r}=T_{c}/T_{0}$, and $\Theta$ is defined in expression
(\ref{CTE}). In the case of a harmonic oscillator with
$\lambda=16\pi\hbar^{2}c^{2}\kappa^{2} a$, and
$\alpha=1/2(\omega_{0}/c)^{2}$ expression (\ref{REL}) becomes
\begin{equation}
\label{HO}
T^{SB}_{c}=\Bigl(1-T_{r}\Bigr)^{-1}\frac{mc\kappa^{-1}}{\hbar
k_{B}}\Bigl(\frac{5.2}{8\pi a_{ho}}\Bigr)^{1/2}N^{1/12}.
\end{equation}
where we used the usual definition for the characteristic length of
the oscillator
\begin{equation}
a_{ho}=\Bigl(\frac{\hbar}{m \omega_{0}}\Bigr)^{1/2}.
\end{equation}
Clearly, the relation between $T_{c}^{SB}$ and $T_{c}$ depends on
the scale of the system $\kappa$. To estimate the order of magnitude
of the scale $\kappa$, let us resort to the definition of the
healing length $\xi$ \cite{Pethick}.
Finally, let us analyze the definition for the chemical potential $\mu$ at the critical temperature.
In the description given above, the relation between the critical temperature and the critical density in the center of the trap (\ref{PQ}) remains the same as for the noninteracting model  assuming that the thermal bath contribution is the lowest energy associated to the system, and it is interesting to look for effects which violate this relation, this work is in preparation \cite{ET}.

\section{Estimation of the Scale $\kappa$}
The healing length $\xi$ is a crossover between the phonon spectrum
and the single-particle spectrum of the Bogoliubov excitations
\cite{AAA}. The healing length is related to the chemical potential
in the center of the condensate and below the condensation
temperature through the next expression \cite{Pethick}
\begin{equation}
\label{HL} \frac{\hbar^{2}}{2m\xi^{2}}=\mu.
\end{equation}
In our case, $\mu$ below the condensation temperature is given by
(\ref{PQ})
\begin{equation}
\mu=\frac{\lambda\kappa^{-2}}{2mc^{2}}n(\vec{r}=0)+\frac{\lambda}{4mc^{2}}(k_{B}T_{c})^{2},
\end{equation}
being $n(\vec{r}=0)$ the density in the center of the condensate.
Using the relation between the healing length and the chemical
potential, allows us to obtain the next expression
\begin{equation}
\label{HL1} \xi=\frac{\hbar
c}{\sqrt{\lambda\Bigl(\kappa^{-2}n+(k_{B}T_{c})^{2}/2\Bigr)}}.
\end{equation}
Resorting to the relation between $\lambda$ and the scattering
length $a$, $\lambda=16\pi\hbar^{2}c^{2}\kappa^{2} a$, we obtain for
the scale $\kappa$,
\begin{equation}
\label{kappa} \kappa=\sqrt{\frac{\xi^{-2}-16 \pi a n}{8 \pi a
(k_{B}T_{c})^{2}}}.
\end{equation}
The order of magnitude of the healing length in typical experiments
is approximately (or bigger than) one micrometer \cite{pana}. In the
case of $^{87}Rb$,  $\xi\sim \hspace{0.01cm} 0.4\hspace{0.09cm}\mu
m$,
 and the scattering length $a\approx 5.77 cm$.
Assuming $T_{c}\approx 200\times 10^{-9}\hspace{0.045cm}K$ say,
together with values of the density $n$ given approximately by
$10^{13}$ to $10^{15}$ atoms per $cm^{-3}$ \cite{Dalfovoro}, allows
us to infer the order of magnitude for the scale $\kappa$ in
ordinary units
\begin{equation}
\label{OM} \kappa \approx 3\times 10^{39} Joules^{-1} Meters^{-3/2}.
\end{equation}

\section{Conclusions}

Starting from the Klein--Gordon equation and the mechanism given in
quantum field theory for the break down of the symmetry of the
system, we where able to reduce this Klein--Gordon equation to a
generalized Gross-Pitaevskii equation (see \cite{Matos:2011kn}). The
question that arises is whether the critical temperature for the
break down of the symmetry of the system and the critical
temperature of the condensation are somehow related or whether they
are of the same order of magnitude. In this work we have given an
answer. We have analyzed the relation between the critical
temperature associated to the spontaneous symmetry breaking of
scalar fields in a thermal bath characterized by a Klein--Gordon
equation and the corresponding condensation temperature. We obtain
the semiclassical energy spectrum associated to this system, from
which we deduced the spatial density, interpreting $\kappa ^{2}$ as
the scale that relates the usual definition of spatial density
within the semiclassical approximation and the density associated to
the self--interacting part of the potential in the Klein--Gordon
equation. We show that the condensation temperature is independent
of the thermal bath, with the aforementioned conditions, and we
prove that $T^{SB}_{c}>T_{0}>T_{c}$, when the coupling constant is
positive. Additionally, we estimate the order of magnitude of the
scale $\kappa$, which according to the definition (\ref{FE}) and the
relation (\ref{kappa}), is very small in typical experiments. These
arguments shows us that, in principle, the relevant thermodynamic
properties associated to the system, in particular the healing
length $\xi$ , could be used to provide  bounds for the scale
$\kappa$, and consequently for $T_{c}^{SB}$ and its relation with
the condensation temperature.
\begin{acknowledgements}
This work was partially supported by CONACyT M\'exico under grants
CB-2009-01, no. 132400, CB-2011, no. 166212,  and I0101/131/07
C-234/07 of the Instituto Avanzado de Cosmolog\'ia (IAC)
collaboration (http://www.iac.edu.mx/).
\end{acknowledgements}

\end{document}